# Streak formation in flow over Biomimetic Fish Scale Arrays

Muthukumar Muthuramalingam[1], Leo S. Villemin[1,2], Christoph Bruecker[1]
[1]School of Mathematics, Computer Science and Engineering City,
University of London
London, United Kingdom EC1V 0HB

[2]Former Student, School of Life Sciences
University of Keele
Staffordshire, United Kingdom ST5 5BJ

**ABSTRACT**
The surface topology of the scale pattern from the European Sea Bass (*Dicentrarchus labrax*) was measured using a digital microscope and geometrically reconstructed using Computer Assisted Design modelling. Numerical flow simulations and experiments with a physical model of the surface pattern in a flow channel mimic the flow over the fish surface with a laminar boundary layer. The scale array produces regular rows of alternating, streamwise low-speed and high-speed streaks inside the boundary layer close to the surface, with maximum velocity difference of about 9%. Low-velocity streaks are formed in the central region of the scales whereas the high-velocity streaks originated in the overlapping region between the scales. Thus, those flow patterns are linked to the arrangement and the size of the overlapping scales within the array. Because of the velocity streaks, total drag reduction is found when the scale height is small relative to the boundary layer thickness, i.e. less than 10%. Flow simulations results were compared with surface oil-flow visualisations on the physical model of the surface placed in a flow channel. The results show an excellent agreement in the size and arrangement of the streaky structures. From comparison to recent literature about micro-roughness effects on laminar boundary layer flows it is hypothesized that the fish scales could delay transition which would further reduce the drag.

**INTRODUCTION**
All bodies, which move through a surrounding fluid, will generate a boundary layer over its surface because of the no-slip condition at the wall (Schlichting and Gersten, 2017). This boundary layer is a region of concentrated vorticity, which shears the fluid near the body surface and the work done to shear the fluid is the measure of the energy which is spent in locomotion (Anderson et al., 2001). The shear stress near the surface depends on the velocity gradient at the wall and the type of boundary layer, which exist near the surface (Schlichting and Gersten, 2017). If the boundary layer is laminar, the drag will be lesser, but it is more prone to separation at adverse pressure gradients, which increases the pressure drag. A turbulent boundary layer produces more skin friction because of the additional turbulent stress near the surface, however, it can sustain much stronger adverse pressure gradient which allows operating on off-design conditions (Schlichting and Gersten, 2017). There is always a trade-off in design to maintain the initial boundary layer laminar for the maximum extent so that the skin friction drag is lesser (Selig et al., 1995) and changing quickly to turbulent boundary layers in areas which are prone to separation. For marine vehicles, one may overcome larger friction by modifying the surface with a hydrophobic coating so that the fluid slips along the surface in contrast to the no-slip condition of an uncoated one. As a consequence, the skin friction reduces which in turn reduces the net drag of the body (Ou et al., 2004; Daniello et al., 2009). This technology was motivated by the lotus-effect, reviewed recently in (Bhushan and Jung, 2006). This phenomenon helps in even self-cleaning of the dirt on the surface which could reduce fouling in the marine environment (Bhushan et al., 2009). For large fast aquatic swimmer such as sharks, there has been numerous experimental and computational studies on the skin denticles (Wen et al., 2014; Oeffner and Lauder, 2012; Domel et al., 2018). Those were found to manipulate the near skin flow to reduce turbulent drag. However, there is very little work on smaller and slower fish with laminar or transitional boundary layer and the role of different arrangements and patterns of fish scales on their swimming behaviour and hydrodynamics. Up to now, there are only hypotheses about the role of fish scale in hydrodynamics, reported in a recent article by (Lauder et al., 2016) who claimed also that there is still no detailed proof of their hydrodynamic function. The scale morphology of bluegill sunfish was measured successfully with GelSight technology and hinted about the possible hydrodynamic uses of the scales (Wainwright and Lauder, 2016). Later, using the same technology the surface topography of various fish species was measured with and without the mucus layer (Wainwright et al., 2017). Some physical characteristics of scales from grass carp (*Ctenopharyngodon idellus*) were measured and manufactured as a bionic surface and the first indication of drag reduction of about 3% was reported (Wu et al., 2018). They claimed a water-trapping mechanism to be responsible for this reduction, mainly due to flow separation behind the scales. No further details were given on the flow structure. In addition, the scales were not overlapping but treated as individual elements. The present paper aims to reproduce the fish surface in a more realistic way based on statistics of scale measurements and reproduction of the overlapping scale array along the body. We focus our studies on the European Bass, which is a fish commonly found in Mediterranean, North African and North Atlantic coastal water regions. The fish scale pattern and array overlap are almost homogeneous over the length of the body.

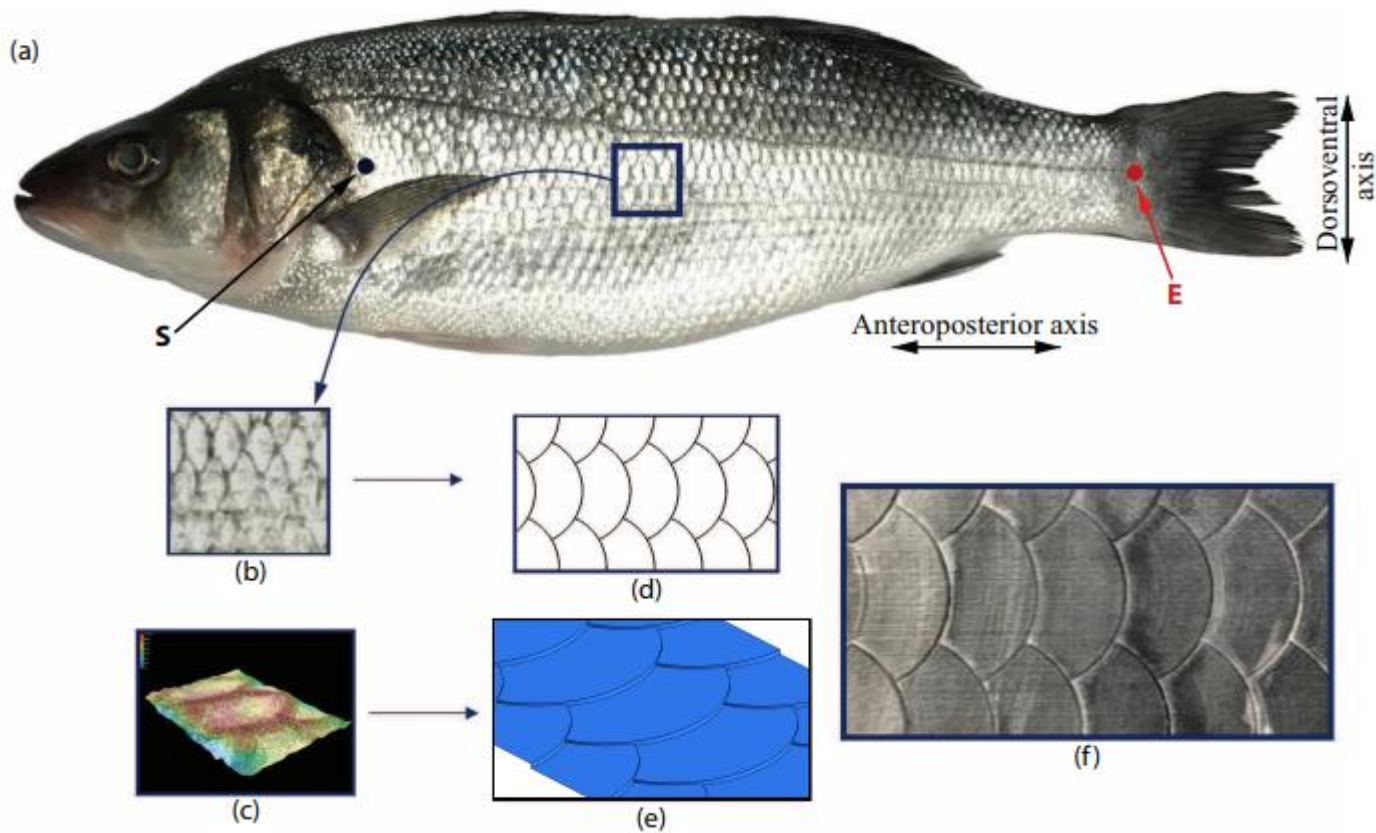

Figure 1: **Microscope images and CAD replication of Sea bass fish scales** (a) Picture of full Sea bass fish. Point S and E represents the region from anterior to posterior where we took measurements of the scales. (b) Top view of the scales (c) Topographical view from scanning with the Digital Microscope (d) Top view of replicated CAD model (e) Isometric view of CAD model (f) Photograph of 3-D printed model from top.

**MATERIALS AND METHODS Fish Samples**

European bass was collected from a local fishmonger (Moxon's Fishmonger – Islington, London). In total, five individuals were analysed from both sexes and had a total length of 33 cm and above. Sampling occurred from the pectoral region to the caudal region at ten equally spaced intervals between point S and E as shown in Fig.1a. The skin of the fish was cleaned repeatedly with a 70% ethanol solution to remove the mucus layer in order to analyse the scale surface under the microscope. Immediately after cleaning, the scale samples were removed from the skin and placed on object slides. The samples were then analysed with a digital microscope (VHX-700FE series, Keyence) using the 3D mapping feature of the built-in software. This allowed to scan the 3D contour and to store the coordinates for later replication of the scale surface in Computer Aided Design (CAD) software. The 2D images and the 3D topographical scan from the microscope, the replicated CAD design and the 3D printed surface of fish scale array are shown in Fig.1 b-f. The physical model was scaled 10 times larger than the actual size, a practical scale for experimental studies in the flow channel. Experiments with up or down-scaled models are a common strategy in hydrodynamic and aerodynamic research based on the boundary layer scaling laws explained in Appendix 2.

**Computational Methodology**

The computational domain and the boundary conditions are shown in Fig.2a. For comparison with the experiments, the length-scale of reference herein is the same as for the 10-times up-scaled physical model. The dimension in 'x' (anteroposterior axis), 'y' (dorsoventral axis) and 'z' (lateral axis) directions are 250mm, 200mm and 80mm. The array of scales is designed with 10 rows along 'x' direction and 5 rows in the 'y' direction. The scale height from the base varies both in 'x' and 'y' direction. Hence, the height of the scale at a given position P(x,y) is defined as $h(P)$, whereas the maximum height of the scale in the centreline ($h_s$) is about 1mm, which corresponds to a 10-times enlarged value compared to the measured value of 100 microns. At the inlet to the domain, a laminar Blasius-type boundary layer velocity profile with a boundary layer thickness ($\delta$) of 10mm was imposed. This profile can be approximated according to Pohlhausen (Panton, 2013) as a second order polynomial profile given by the Eqn.1.

$$\frac{u(y)}{U_\infty} = A\left(\frac{y}{\delta}\right) + B\left(\frac{y}{\delta}\right)^2 \qquad (1)$$

$$\delta(x) = \frac{5 \cdot x}{\sqrt{Re_x}} \qquad (2)$$

$$Re_x = \frac{\rho \cdot U_\infty \cdot x}{\mu} \qquad (3)$$

where A and B are the coefficients based on the free stream velocity ($U_\infty$ = 0.1$ms^{-1}$) and the boundary layer thickness ($\delta$ = 10mm) at the inlet (A = 2, B = −1). The boundary layer thickness given by Eqn.2 corresponds to a flat plate Reynolds number of about $Re_{xo}$ = 33000 with an imaginary inlet length of $x_o$ = 333mm from the leading edge of a flat plate until it reaches the inlet of the domain, where the Reynolds number ($Re_x$) is defined by Eqn.3. Except for the floor and the fish scale array, all the other side walls were specified with free slip conditions, i.e. zero wall-shear. The domain was meshed with 18 million tetrahedral elements with 10 prism layers near the wall with a first cell value of 0.06mm. To study the effect of the scale height relative to the boundary layer thickness on total drag, different boundary layer thickness at the entrance were simulated. Therefore, the inlet domain was extended for 200mm upstream as shown in Fig.2b and the boundary layer thickness at the new inlet was specified as 5,10 and 15mm. The problem was solved using the steady state pressure based laminar solver in ANSYS Fluent 19.0 with a second-order upwind method for momentum equation. Water was used as the continuum fluid in this CFD study with a density ($\rho$) of 1000 $kgm^{-3}$ and a dynamic viscosity ($\mu$) of 0.001 $kgm^{-1}s^{-1}$.

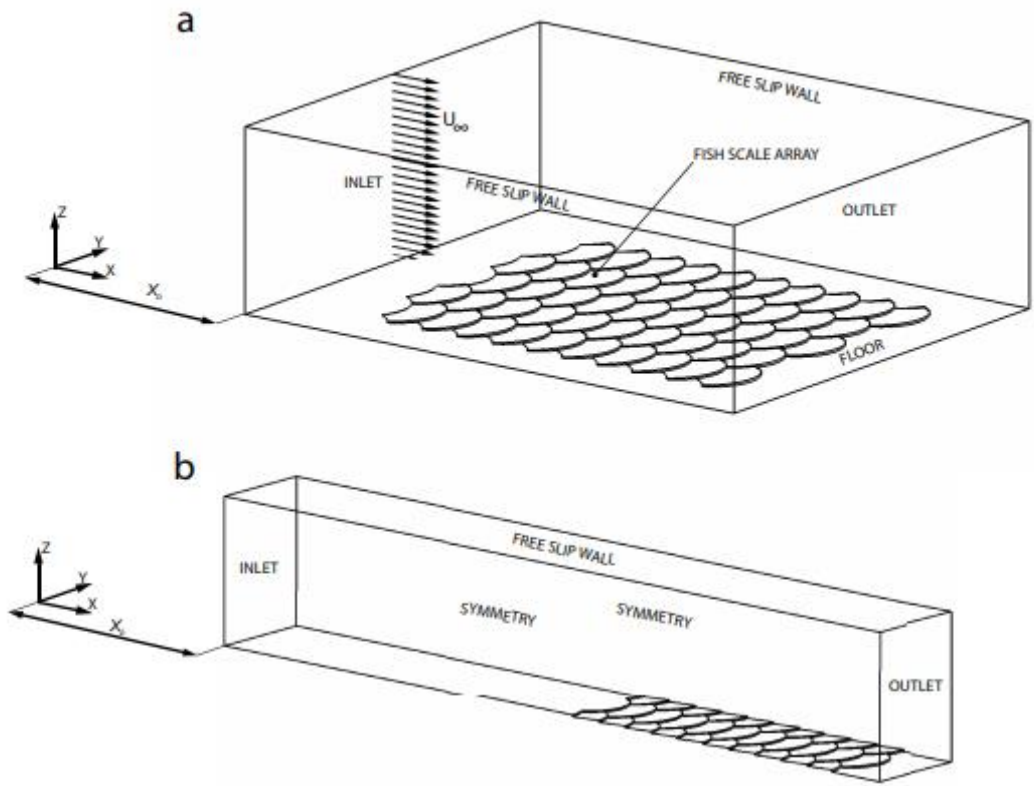

Figure 2: **Computational domain with boundary conditions.** (a) Configuration of the fish-scale array in CFD similar to the condition of the physical model of the scale array at the bottom wall of the wind tunnel. Note that the velocity vector represents the inlet profile with free-stream velocity parallel to the 'x' axis in positive direction (Anteroposterior direction). 'y' axis represents the spanwise (Dorsoventral direction) and 'z' axis represents wall normal direction. (b) CFD domain with symmetry conditions to simulate the drag variation with no end effects. In both the figures '$x_o$' is the imaginary length from the leading edge of the plate to the inlet of the domain.

**Surface Flow Visualisation**

The fish scale array with dimensions explained in the previous section was 3D printed with ABS plastic using Fused Deposition Modeling (FDM)(Printing machine - Raise 3D). For manufacturing, the base layer thickness needed to be 4mm to ensure stable handling. The model was placed on the floor of a wind tunnel (PARK Research Centre, Coimbatore, India) in the test section (cross-section of 450mm and 600mm width). To reduce the disturbance of the step at the leading edge, a chamfered flat plate (size 250mm x 200mm x 4mm) was placed upstream and downstream such that the region with the scale array is flush with the wall. Surface oil-flow visualisation was performed with a mixture of Titanium-di-oxide, kerosene and a drop of soap oil added to it to avoid the clustering of particles. More details of this visualization can be found in (Merzkirch, 2012). Before starting the wind tunnel, the model was painted with the mixture in the region downstream to the scale array. Thereafter, the tunnel flow was started to a free-stream velocity of 12$ms^{-1}$ which gave a boundary layer thickness of about 10mm at the entrance to the scales. Wind is transporting the dye according to the local wall shear and a camera mounted on the top of the tunnel is capturing this process.

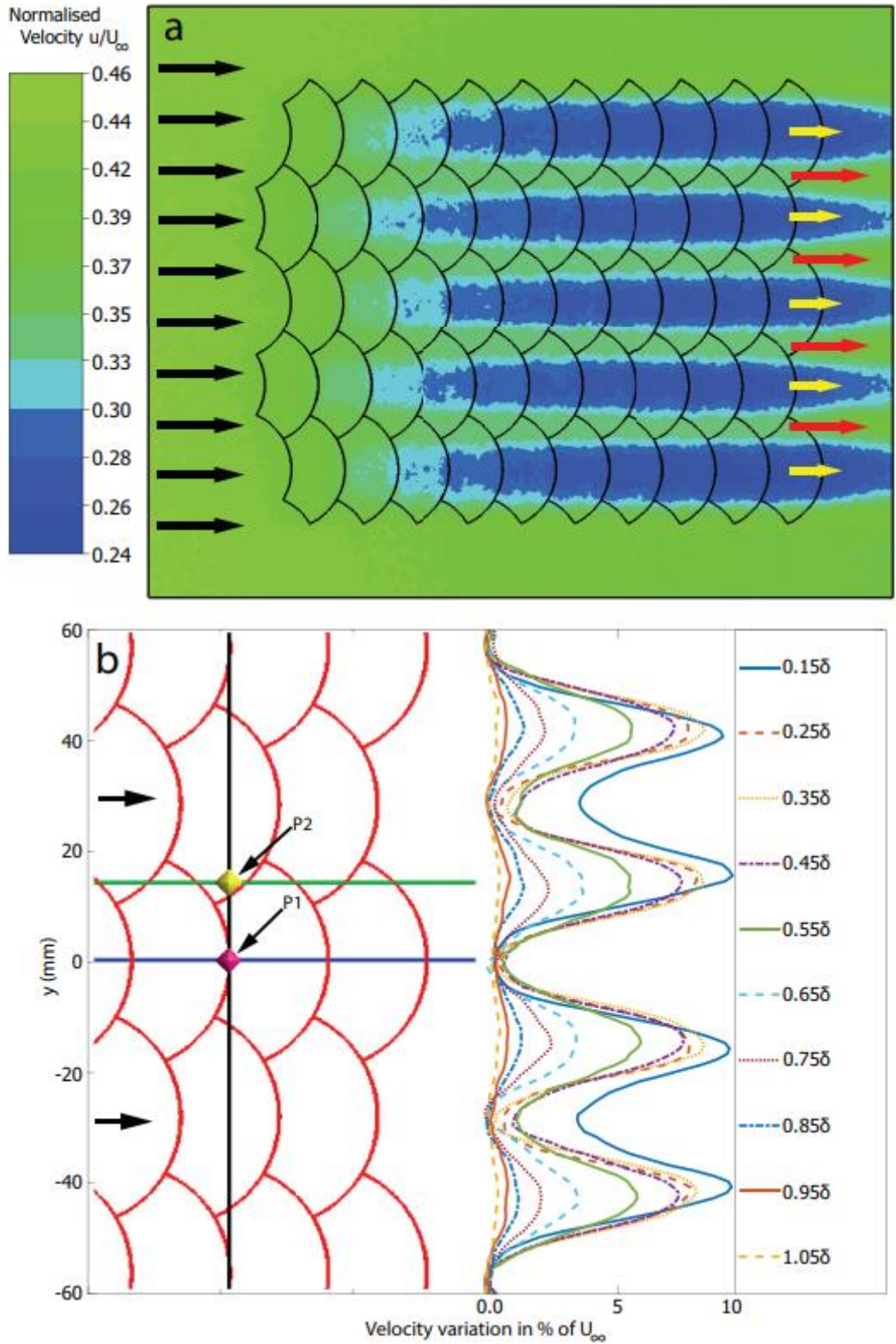

Figure 3: **Velocity contour and velocity profiles** (a) Normalised Velocity Contour at a wall-parallel plane at a distance of z = $0.25\delta$ from the surface. The arrows indicate the flow direction. Note that the black arrows at the inlet are uniform in length, while red and yellow arrows at the outlet differ in length. (b) Velocity variation in spanwise direction at various wall-normal distances in the boundary layer. Scale array is shown in red color for better illustration. **Blue line (Line-1)** represents a centreline of a row of scales. **Green line (Line-2)** represents the overlap region between the scales. Black line represents a location in the 'x' direction at 190mm from inlet. Location1 (P1) and Location2 (P2) are probe points at 190mm from inlet on centreline region and overlap region. Black arrow indicates mean flow direction

## Results

Flow data obtained from the CFD results are first presented as velocity fields and profiles. Fig.3a shows colour-coded contours of constant streamwise velocity (normalised with the free-stream velocity) in a wall-parallel plane at a distance of $0.25\delta$. At the inlet, the velocity is uniform along the spanwise direction ('y' direction), whereas, along the flow direction over the scales, there is a periodic velocity variation in spanwise direction. Low-velocity regions have emerged in direction of the centrelines of the scales, which is indicated with yellow arrows. In comparison, high-velocity region (Red Arrow regions) are seen along the regions where the scales overlap each other. These high velocity and low-velocity regions are referred in the following as streaks. These structures are linked in number, location and size with the overlap regions along the dorsoventral axis over the surface.

Further information of the variation of the velocity in the streaks is demonstrated Fig.3b. It shows spanwise profiles of the streamwise velocity at the location $x = x_o + 190mm$ (8$^{th}$ scale in the row along streamwise direction from the inlet) for different wall normal locations. At a wall normal location of $0.15\delta$ the velocity variation is around 10% of $U_\infty$ between the peak (local max) and valley (local min) in the profile. Given this difference, the streak amplitude is calculated using the Eqn.4 from (Siconolfi et al., 2015).

$$A_{ST} = \left[\max_y\{U(X,y,z)\} - \min_y\{U(X,y,z)\}\right] \Big/ (2U_\infty) \tag{4}$$

As seen from the different profiles, the location of peaks and valleys do not change with wall normal position, therefore the streaks extend over most of the boundary layer thickness in a coherent way. The streak amplitude $A_{ST}$ is plotted along the wall normal location in Fig.4 and it can be seen that the streak amplitude is maximum within the first 20% of the boundary layer thickness with a value of 4.5% of $U_\infty$. As the distance from the wall increases, the streak amplitude decreases monotonically until the displacement effect of the scales has died out at the outer edge of the boundary layer ($1.05\delta$ from the wall).

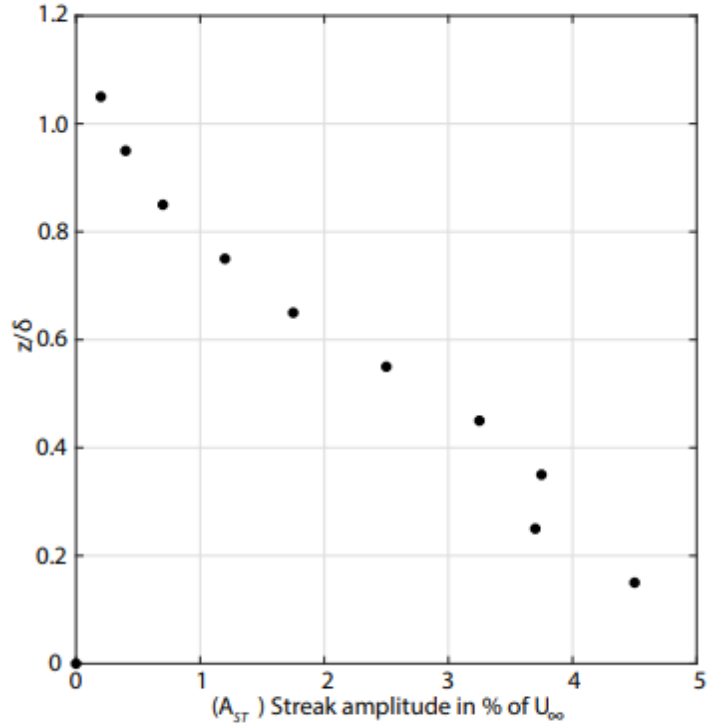

Figure 4: **Variation of streak amplitude along wall-normal direction at 190mm from inlet (refer black line in Fig.3b**

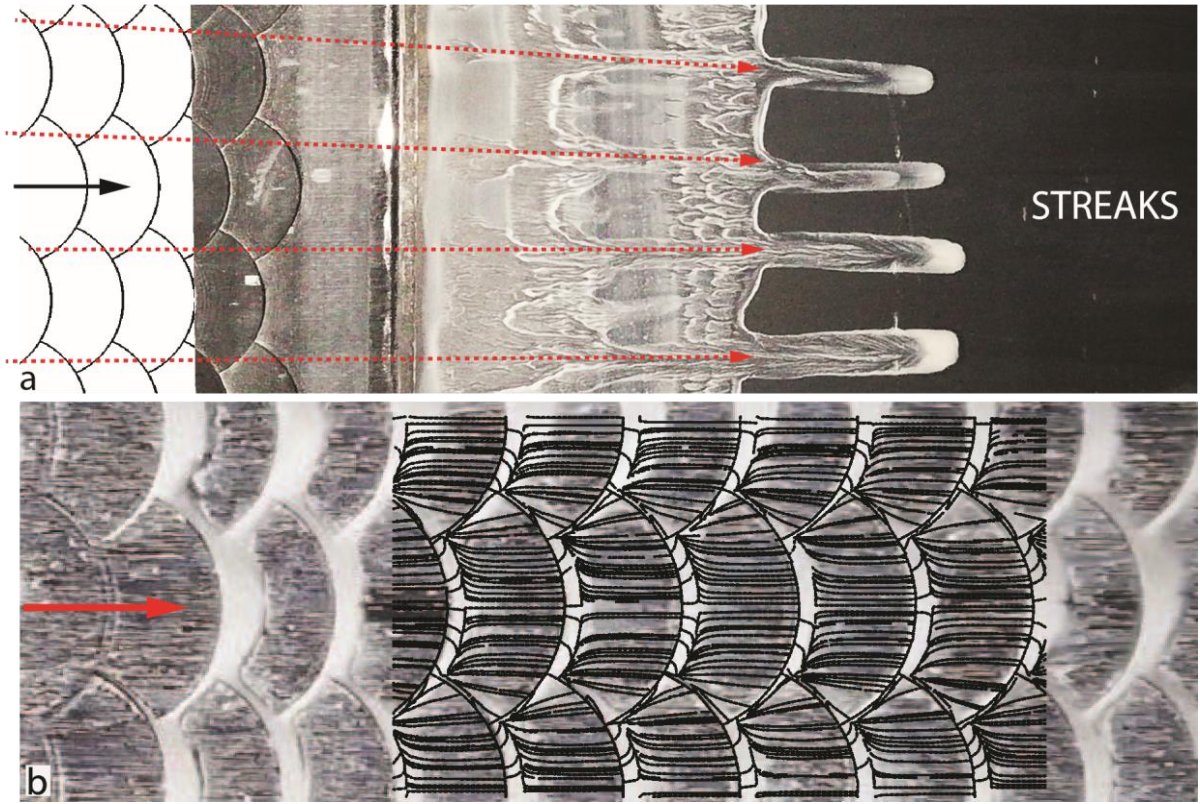

Figure 5: **Surface flow visualisation picture from top view onto the scales.** (a) Black arrow represents mean flow direction. Note that the oil-mixture was painted in the region downstream of the scales to highlight the generation of the streaks. 2-D top view of the CAD model is merged to get the impression of the arrangement of the scales. The red arrows were added to illustrate the trace of the streaks relative to the arrangement of the scales. (b)Red arrow represents the mean flow direction. Herein, the oil-mixture was painted directly onto the scales. Surface streamlines from CFD simulation are overlaid to compare the results. Note that the regions of accumulated oil-patches match with the regions of flow reversals from CFD simulation.

Experimental flow visualisation pictures of the streaks behind the fish scale array are shown in Fig.5a. As the particle mixture coated on the surface moves according to the direction and the magnitude of wall-shear, the mixture moves farther in the regions of high shear, than in regions of low shear. Therefore, the flow produces streaky patterns on the surface with different length downstream of the scale array. This can be observed in Fig.5a from finger-like pattern of which each finger represents a high-speed streak. The red lines depict the orientation of the streaks relative to the pattern of the scale array. It is clearly seen that the high-speed streaks are formed in the overlap regions as previously claimed from the CFD results. For better comparison with the CFD results, the surface flow visualisation over the scale array is overlaid with surface streamlines from CFD (see Fig.5b), which is discussed later.

Figure.6 shows the variation of the normalised velocity profile at two locations along the span at the $8^{th}$ scale row (probe point location P1 and location P2 , compare Fig.3b). In the absolute coordinate system (Fig.6a) there is a shift in 'z' direction because of the variation in the scale height h(P) along the span of the surface. When the profiles are plotted in the body relative system ($z = z - h(P)$) the difference along the wall normal direction (Fig.6b) becomes more obvious. With the scales on the surface, the gradient of the velocity near the wall gets steeper in the location discussed here (at the probe points P1 and P2). This is concluded from the comparison to the theoretical velocity profile for a smooth flat plate (dashed black line, from Eqn.1). However, the boundary layer thickness is approximately the same. Hence, it could be concluded that the scales change the profile shape inside the boundary layer region but would not change the boundary layer thickness (nevertheless affecting the displacement and momentum thickness).

The surface streamline picture generated from the CFD results is shown in Fig.7a. In the centreline of the scales, the flow mostly follows the direction of the main flow. Section X-X is enlarged and the cross-sectional flow in the centre of the scales is shown in Fig.7b. It is seen that the flow follows the small slope caused by the tilt angle of the scale until it separates from the sharp edge on the scales and reattaches further downstream at approximately 2.5 times the scale height ($h_s$) on the surface as a laminar boundary layer. This non-dimensional reattachment length is very similar to the value reported in horizontal backward facing step flows if the Reynolds number defined with the step height and free stream velocity is around 100 for the given flow situation(Goldstein et al., 1970). This separated flow region behind the step is visible from the dividing streamline (shown as thick dotted line in Fig.7b). Also, from the surface flow visualisation, the separated flow region behind the edge of the scales can be observed by the white patches due to the accumulation of the particles (see Fig.5b). These white-patched regions match in size and locations with the flow reversal zones in the CFD. When the fluid moves along the scales, the streamwise component of velocity is reduced in the central region of the scale by the large separated zone as explained above. This causes a spanwise pressure gradient and forces the fluid to move from the central region of the scales to the overlapping region. This movement is seen in the zig-zag pattern (shown in blue arrows in Fig.7a) with larger spanwise components of fluid motion. The spanwise flow towards the overlapping region produces a higher streamwise velocity because of mass conservation. This causes high-speed streaks in these regions. In addition, it is evident that the flow reversal is reduced compared to the cross-section at the central region of the scales. This is the root cause of producing low speed and high-speed streaks.

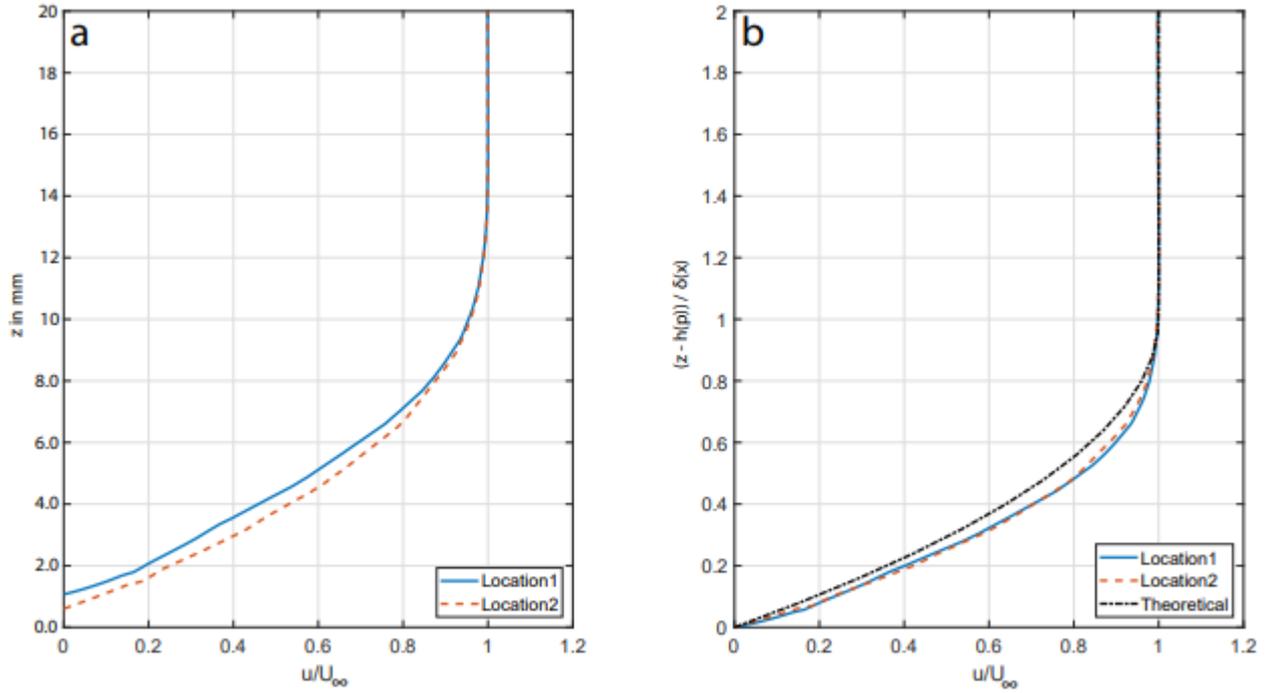

Figure 6: **Boundary layer profiles at Location 1 and Location 2 (probe points P1 and P2 in Fig.3b).** (a)Normalised velocity profiles in the absolute coordinate system. Note that the shift in velocity profiles along 'z' direction is because of the change in scale height h(P) for the different probe points P (b)Normalised velocity in the body coordinate system with theoretical boundary layer profile along a smooth flat plate.

Figure.8 shows the surface streamlines on the scale array along with cores of intense vortices visualised by isosurfaces of the 'Q' value (Jeong and Hussain, 1995). The colours of the isosurfaces indicate the streamwise helicity which is defined as ($U_x.\omega_x$), where, $\omega_x$ is the vorticity component along 'x' direction. The yellow colour region defines the region in which the vortex direction is Counter Clockwise (CCW) with respect to the 'x' axis direction (i.e. mean flow direction represented by a white straight arrow in Fig.8.), similarly, the blue colour region defines the vortex direction in Clockwise (CW). It also displays the cross-flow velocity fields on planes parallel to the Y-Z plane near the scale overlap region for two consecutive scales. The vortex in the central region of the scales (i.e. white colour vortex core) reflects the reversed flow region behind the step. There the flow direction remains nearly aligned with the mean flow. In comparison, when the flow moves downstream in the overlap region it is affected by successive vortices with alternating direction switching from CCW to CW and vice versa. This causes the streamlines in the overlap region to generate a zig-zag pattern as already illustrated in Fig.7a.

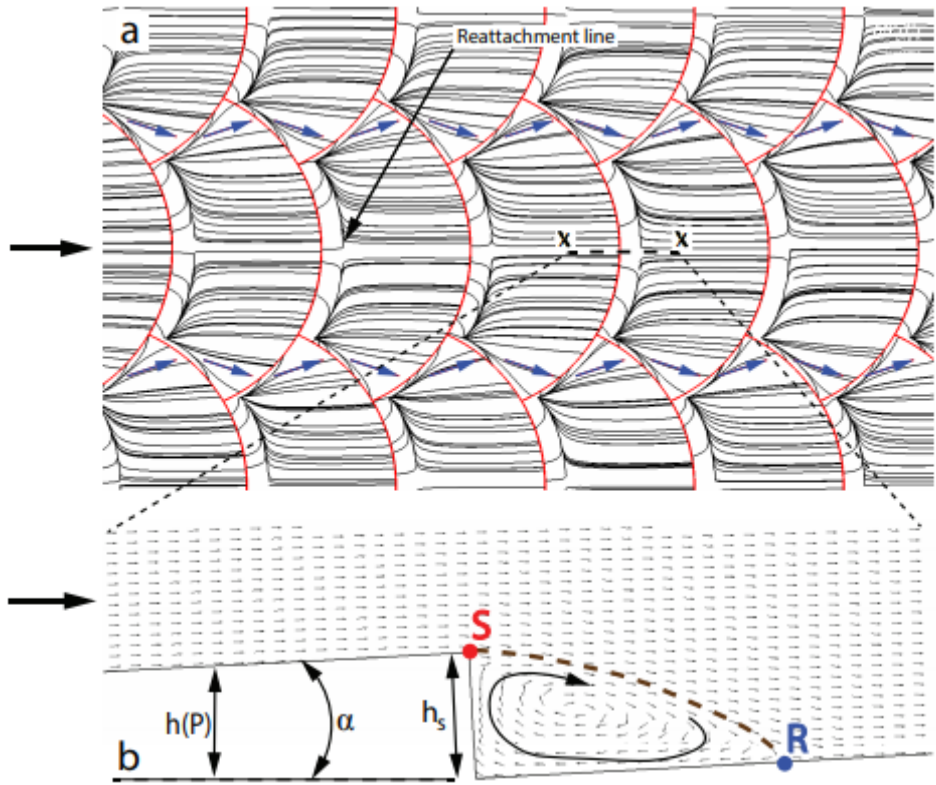

Figure 7: **Surface streamline and vector plots from CFD simulation over the scales.** (a)Top view of surface streamline over the scales. Note the zig-zag motion along the overlap region compared to the parallel flow at the central regions of the scales. (b)Vector field in the x-z cross sectional plane along the central region of the scale at line X-X (vectors indicate only direction and not magnitude). $h_s$ = 1$mm$ and $\alpha$ = 3 degrees. S and R represents the separation and reattachment of the flow streamlines. Thick dashed line indicates the region of recirculating flow with an arrow indicating the direction of rotation. In both drawings the black arrow indicates the mean flow direction.

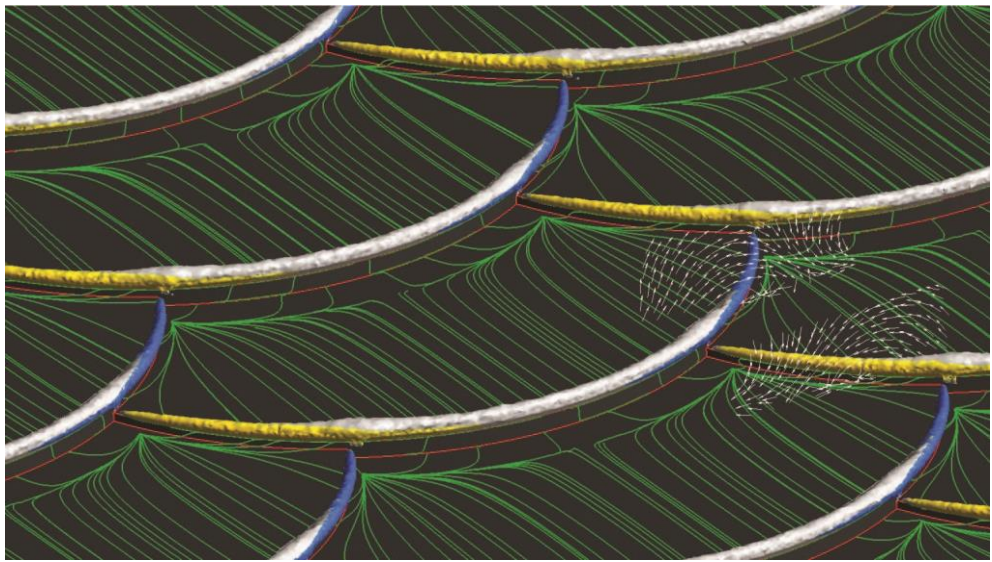

Figure 8: **Surface streamline plot with direction of vortices.** Vector plots at the overlap region for two consecutive scale rows. Helicity coloured in yellow for positive (Vortex direction CCW) and blue for negative (Vortex direction CW). Other rotational vectors are based on the colouring of vortices. Vortices are identified with 'Q' criterion. White straight arrow represents the mean flow direction.

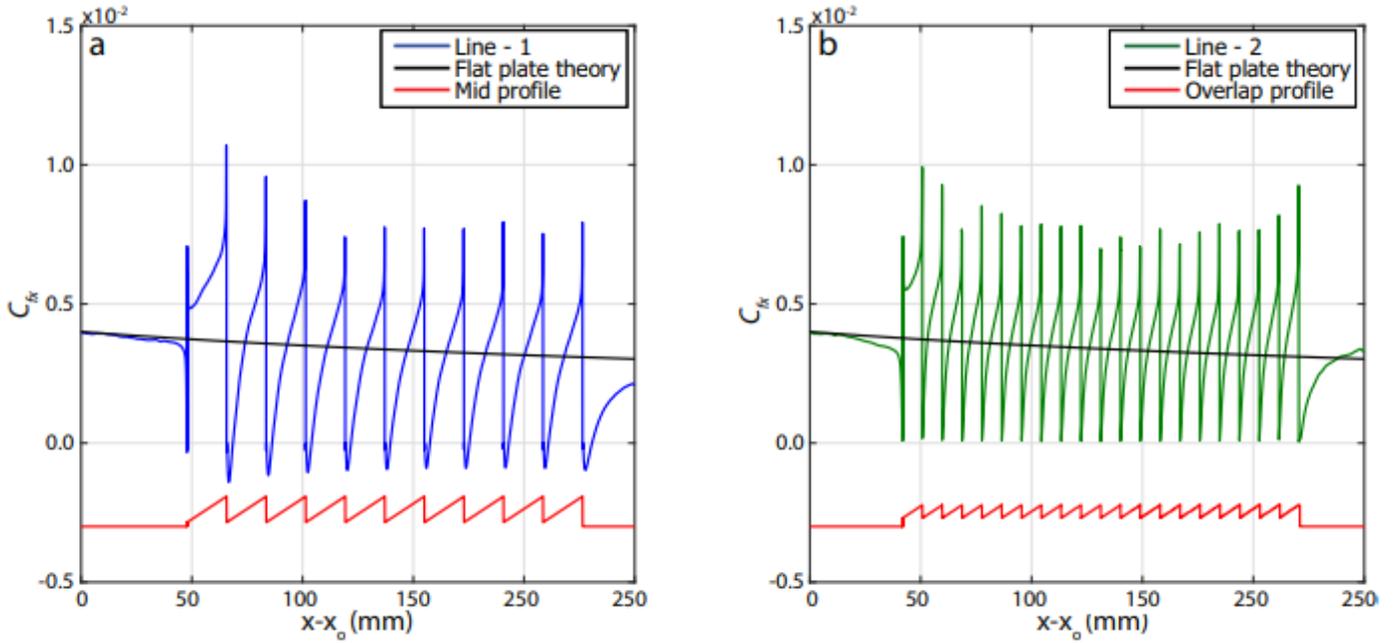

Figure 9: **Variation of skin friction coefficient** ($C_{fx}$) **along 'x' direction at two locations.** (a)Blue line represents the $C_{fx}$ variation along **(Line-1)** (refer **Blue line** in Fig.3b) (b)Green line represents the $C_{fx}$ variation along **(Line-2)** (refer **Green line** in Fig.3b). Red lines represents variaion of 'z' coordinate in 'x' direction along corresponding locations (not to scale). Black line represents the variation of $C_{fx\,theory}$ for flat plate boundary layer by Eqn.5. $x_o$ (333mm) is the imaginary length before the inlet of the domain.

**Skin Friction and Total Drag**

As previously mentioned, the scales modulate the near wall flow with streaks which will change the wall shear stress ($\tau_w$) distribution on the surface when compared with flow over smooth flat plate. To analyse this effect, skin friction coefficient $C_{fx}$ defined by Eqn.5 is plotted along the centreline (see **Blue line (Line-1)** in Fig.3b) together with the surface profile variation in Fig.9a. In addition, the figure shows the profile of the theoretical skin friction coefficient ($C_{fx\,theory}$) for a smooth flat plate case, given in Eqn.6. Along the initial smooth part of the surface until 25mm the skin friction coefficient follows the theoretical skin friction coefficient $C_{fx\,theory}$. As it enters the scale region, initially the skin friction drops because of the adverse pressure gradient caused by the first wedge. Over the scale, it increases again because of the local acceleration until it reaches the maximum at the edge of the scale. Then, $C_{fx}$ drops to a negative value because of the recirculation region explained in Fig.7b. Once the flow reattaches, the skin friction gets positive again and increases until it reaches the peak as it approaches the edge of the next scale. This process repeats itself in flow direction with the succession of scales. The same process happens in the overlap region, but here, for a single scale length, the process happens twice because of two small steps formed by the adjacent scales in the lateral overlap region (note the difference in the scale profile in the central region in Fig.9a and the scale profile in the overlap region in Fig.9b). Additionally, the streamwise wall shear does not reach negative values in the valleys as there is no flow reversal in these zones. The shear drag along the central region (determined by the integration of wall shear in the streamwise direction along **Blue line (Line-1)** in Fig.3b) gives a 12% reduced value compared to the theoretical drag for a smooth flat plate. In contrast, the overlap region (determined by the integration of wall shear in the streamwise direction along **Green line (Line-2)** in Fig.3b) gives a 5% increase in shear drag. This tendency along the span correlates with the low and high-velocity regions as the wall shear stress is directly proportional to the velocity gradient. The integral over the total surface leads to the total friction drag, which is a net effect of the streaks. As we introduce a surface which is not smooth, the total drag is the sum of the friction and the pressure drag. The latter depends on the wake deficit behind the step of the scale because of the separated flow regions. Both need to be taken into account from the CFD results to investigate the net effect on possible total drag reduction.

In order to investigate the relative contributions of friction and pressure drag over the skin, we varied the boundary layer thickness ($\delta$) relative to fish scale height ($h_s$) as reported in Table 1. The inlet boundary layer thickness in the CFD domain was increased in steps from $\delta$ = 5*mm*, 10mm and 15mm respectively with a free stream velocity ($U_\infty$) value of $0.1 ms^{-1}$. Drag coefficients were calculated using the drag force values obtained from CFD. The change in friction drag and total drag coefficients is given in Eqn.7. The theoretical drag coefficient ($C_{d\ theory}$) is calculated by integrating the skin friction coefficient ($C_{fx\ theory}$) along the 'x' direction.

$$C_{fx} = \frac{\tau_w}{0.5 \cdot \rho \cdot U_\infty^2} \tag{5}$$

$$C_{fx\ theory} = \frac{0.73}{\sqrt{Re_x}} \tag{6}$$

$$\Delta C_{df}(\%) = \frac{(C_{df} - C_{d\ theory})}{C_{d\ theory}} \times 100 \quad \Delta C_{d\ tot}(\%) = \frac{(C_{d\ tot} - C_{d\ theory})}{C_{d\ theory}} \times 100 \tag{7}$$

Table 1: **Dependence of drag force with boundary layer thickness to fish scale height ratio**

| $\delta/h_s$ | $C_{dp}$ | $C_{df}$ | $C_{dtot}$ | $C_{dtheory}$ | $\Delta C_{df}$ (%) | $\Delta C_{dtot}$ (%) |
|---|---|---|---|---|---|---|
| 5 | 0.000277 | 0.00448 | 0.00476 | 0.00453 | -1.03 | 5.08 |
| 10 | 0.000193 | 0.00301 | 0.00320 | 0.00316 | -4.68 | 1.43 |
| 15 | 0.000129 | 0.00214 | 0.00226 | 0.00236 | -9.31 | -3.84 |

For all the three cases the change in friction drag ($\Delta C_{df}(\%)$) relative to the smooth flat plate is negative indicating that the scales are efficient in reducing skin friction. This effect increases with increasing boundary layer thickness to scale height ratio. However, the total drag is only reduced for the third case ($\Delta C_{d\ tot}$ = −3.84%) when $\delta/h_s$ ratio is 15. This is the typical ratio between the boundary layer thickness and the scale height in cruising conditions of the flow around the fish and will be explained in the discussion.

**DISCUSSION**

In this paper, 3D microscopic measurements of the scales on the European bass fish are presented. Based on the data statistics, a biomimetic scale array was replicated with the use of Computer Aided Design and 3D printing. The study differs from previous ones on biomimetic scales (Dou et al., 2012; Wainwright et al., 2017; Wainwright and Lauder, 2017) that it is the first for European bass and the first using a typical 3D curvature of the scales with an additional overlap pattern. Flow over the scale array was analysed using Computational Fluid Dynamics and experimental results were obtained from the surface flow visualisation. Excellent qualitative agreement was found, showing the formation of alternating high-speed and low-speed streaks along the span, which concludes that the location, size and arrangement of the streaks are linked with the overlap pattern of the scales. The experimentally validated CFD data further allows drawing conclusions about the total drag of the surface, which is relatively difficult to obtain. The derived drag values show that the overlapping scale arrays are able to reduce the body drag if their characteristic step height is sufficiently small (at least one order of magnitude) compared to the local boundary layer thickness. If this conclusion holds for typical flow conditions and size of the scale for European bass, the consequence would be a reduction of total drag, hence costing less energy to the fish in cruising. In the following, we discuss the possible relevance of this finding to the situation of sea bass in steady swimming conditions, including a critical review of the limitations of the study.

**Mucus layer and transport:**
- Any mucus on the scales needed to be washed away for optical reasons before the scales could be measured in the microscopy. It is known for similar fish species that the mucus only covers the microstructures of the scales such as circulae and the ridges which connect the ctenii, therefore the overall shape of the scales is not affected by the wash-out procedure (see also the conclusion by (Wainwright et al., 2017)). Thus the flow dynamics is representative for the natural situation of the scales in the flow.
- The observed recirculating flow near the central region of the scales might be helpful in retaining the mucus and reducing the mucus secretion rate. This inference is supported from the fact that in the surface flow visualisation experiments the mixture was largely trapped in these regions. This is comparable with the results on flow over grass carp fish scales(Wu et al., 2018).

**Swimming speed and Reynolds number:**

- The swimming speed of European bass is proportional to its body length (Carbonara et al., 2006). For the fishes considered in this study, the swimming speed lies in the range from 1.2 $ms^{-1}$ and 1.4 $ms^{-1}$ corresponding to a Reynolds number (calculated with the full body length L) in the range between $4 \times 10^5$ and $6 \times 10^5$. This is in classical fluid dynamics when transition from laminar to turbulent boundary layer flow sets in. As the reference length is the tail end, we can conclude that the boundary layer over the sea bass for most of the body length remains laminar. Direct measurements of the boundary layer on sea bass are not known so far, however, such data exist for comparable fish such as a scup (*Stenotomus chrysops*), a carangiform swimmer, and rainbow trouts (*Oncorhynchus mykiss*). Scup have mostly an attached laminar boundary layer over its body for most of the time and incipient separation appears only for short time intervals in the swimming cycle (Anderson et al., 2001). PIV analysis on swimming rainbow trout at a Reynolds number of $4 \times 10^5$ revealed a laminar boundary layer with transition to turbulence in the caudal region (Yanase and Saarenrinne, 2015). Hence, the laminar CFD analysis performed in this study is representative for the effect of fish scales on typical European bass.

- For the total drag of the biomimetic surface, a drag reduction was only observed when the scale step-height was sufficiently small relative to the local boundary layer thickness (one order of magnitude). At a swimming speed of 1.2 $ms^{-1}$ and for a fish length of 300mm the boundary layer thickness is 1.5mm at the mid of the fish body (from Eqn.2) measured from the snout of the fish to the begin of the caudal fin. In this region, the scale height measured from the microscope was about 0.1mm which gives a boundary layer thickness to scale height ratio ($\delta/h_s$) of 15 and has proven reduction in drag. Interestingly, the boundary layer thickness of scup is also in the same range discussed here. Hence, the study shows, at least for steady swimming conditions, valid implications on total drag reduction due to the presence of overlapping scale arrays.

**Relevance of streaks in boundary layer transition:**

- In studies of a boundary layer flow over a flat plate it has been shown that placing arrays of micro-roughness elements on the plate can delay transition (Fransson and Talamelli, 2012; Siconolfi et al., 2015). The effect of those elements is that they produce low speed and high-speed streaks inside the laminar boundary layer, which delay the non-linear growth of the Tollmien–Schlichting waves (Fransson et al., 2004). Although the mechanisms to generate the streaky pattern might be different (lift-up mechanism of streamwise vortices versus alternating vortices in the overlap regions), the fish-scale array producing streaks could also lead to the delay in transition.

To summarize, the biomimetic fish scale array produces steady low and high-speed streaks, which are arranged in spanwise direction in the same pattern as the rows of the overlapping scales. Those regular arrangement of streaky structures are known from flow studies on generic boundary layer flows to stabilize the laminar steady state and delay transition to turbulence. As already mentioned, the swimming Reynolds number of the fish considered here lies in the transitional range. Thus, we conclude that steady streaks similar as those observed for the biomimetic scale array are indeed produced by the scales of fish and help to maintain laminar flow over the fish body. The presented biomimetic surfaces can be engineered by purpose to reduce skin friction and delay transition in engineering application. However, this only refers to steady swimming conditions. Undulatory motion of the body during active propulsion plays an additional role in the boundary layer transition. Experiments with undulatory moving silicone wall in flow show an alternating cycle between re-laminarization and transition in the trough and at the crest of the body wave (Kunze and Bruecker, 2011). As the fish surface can also undergo bending motion, the overlapping scales can move relative to each other and deploy in regions of strong curvature. From previous measurements of the boundary layer over swimming scup, it is known that the boundary layer remains laminar for most of the body without flow separation even in the adverse pressure gradient region (i.e. aft part of the fish) (Anderson et al., 2001). If the scales therefore also take part in any manipulation of flow separation is still an open question (Duriez et al., 2006). From a technological perspective, artificial surfaces with scales can even be built from flexible material, addressing also the issue of local flow separation.

**Acknowledgements**

The position of Professor Christoph Bruecker is co-funded by BAE SYSTEMS and the Royal Academy of Engineering (Research Chair No. RCSRF1617\4\11, which is gratefully acknowledged. The position of MSc Muthukumar Muthuramalingam was funded by the Deutsche Forschungsgemeinschaft in the DFG project BR 1494/32-1, which largely supported the work described herein. We would like to thank Mr. Avin Alexander Jesudoss for providing help on performing surface flow visualisation on the fish scale model.